# Tailoring multilayer quantum wells for spin devices

S. ULLAH[1,*], G. M. GUSEV[1], A. K. BAKAROV[2] and F. G. G. HERNANDEZ[1]

[1] Instituto de Física, Universidade de São Paulo, Caixa Postal 66318, CEP 05315-970 São Paulo, SP, Brazil
[2] Institute of Semiconductor Physics and Novosibirsk State University, Novosibirsk 630090, Russia
*Corresponding author. E-mail: saeedullah.phy@gmail.com

**Abstract.** The electron spin dynamics in multilayer GaAs/AlGaAs quantum wells, containing high-mobility dense two-dimensional electron gases, have been studied using time-resolved Kerr rotation and resonant spin amplification techniques. The electron spin dynamics was regulated through the wave function engineering and quantum confinement in multilayer quantum wells. We observed the spin coherence with a remarkably long dephasing time $T_2^* > 13$ ns for the structure doped beyond metal-insulator transition. Dyakonov-Perel spin relaxation mechanism, as well as the inhomogeneity of electron *g*-factor, was suggested as the major limiting factors for the spin coherence time. In the metallic regime, we found that the electron-electron collisions become dominant over microscopic scattering on the electron spin relaxation with the Dyakonov-Perel mechanism. Furthermore, the data analysis indicated that in our structure, due to the spin relaxation anisotropy, Dyakonov-Perel spin relaxation mechanism is efficient for the spins oriented in-plane and suppressed along the quantum well growth direction resulting in the enhancement of $T_2^*$. Our findings, namely, long-lived spin coherence persisting up to about room temperature, spin polarization decay time with and without a magnetic field, the spin-orbit field, single electron relaxation time, transport scattering time, and the electron-electron Coulomb scattering time highlight the attractiveness of *n*-doped multilayer systems for spin devices.

**Keywords.** Spin coherence, Quantum wells, *g*-factor, Spin dephasing, Kerr rotation, Spin-orbit field.

**PACS Nos** 78.20.Ls; 85.70.Sq; 75.25.-j; 76.60.Es

## 1. Introduction

In recent years, the spin dynamics in semiconductor nanostructures has become the focus of intense research due to the possibility of using the spin degree of freedom in future technology [1]. Among the key requirements, for successful implementation of novel spintronic devices, quantum computation, and quantum information processing [2, 3, 4], a suitable system exhibiting low relaxation rate and a long transport length [5, 6, 7] is highly desirable. Those applications could benefit from such systems because they can store and process the information before the decoherence effect set in.

However, due to strong coupling to its environment in a solid-state system, the spins in low-dimensional structures like quantum wells (QWs) and quantum dots (QDs) meet a vital problem of strong dephasing. In this respect, various material structures [8, 9, 10, 11, 12, 13, 14] have been tried to control this fast decoherence. Among those materials, GaAs-based heterostructures have drawn considerable attention because of its numerous properties that make it well suited for applications in telecommunication, high frequency, and high-speed electronics [15].

Recent advances in molecular-beam epitaxy (MBE) enable the engineering of new and advanced multilayer structures. By tailoring the sample geometry, thereby producing the environment to confine the carriers wave function that penetrates into the barriers, one can witness an internal magnetic field (spin-orbit field). Such field is believed to be the tuning force for the spin manipulation [3]. Today, most of the schemes proposed for the generation, manipulation, and detection of spins rely on this internal magnetic field [16, 17, 18]. Recently, the spin-orbit effects have been attracted renewed interest due to the emergence of striking phenomenas such as persistent spin helix (PSH) [19, 20], spin Hall effect [23], large spin relaxation anisotropy [24], and Majorana fermions [21, 22]. Additionally, such wave vector (***k***) driven fields induced by bulk inversion asymmetry (Dresselhaus field) [25] or structure inversion asymmetry (Rashba field) [26] can also inherently result in spin relaxation through DP mechanism [27].

The expectations for device applications of spin-polarized electrons will become more realistic by understanding the microscopic mechanisms responsible for the spin relaxation as well as its manifestation in different experimental conditions, for example, applied magnetic field and sample temperature, etc. It is believed that such relaxation processes are substantially modified in the two-dimensional systems compared to the bulk [2]. While there is a vast literature on the spin relaxation process of electrons in semiconductor QWs, there are only a few investigations of carrier spin relaxation in multilayer structures. In present work, we investigate the electron spin dynamics in multilayer GaAs/AlGaAs structures. Such structures, in principle, allow the long-lived spin polarization as well as the manipulation of those spin through the spin-orbit field [5, 6, 7, 24]. Recently, the authors demonstrated that such multilayer QWs could transport coherently precessing electron spins over about half millimeters at liquid He temperature [6].

The paper is organized as follows. Section 2. presents the material and experimental details. Section 3. is devoted to experimental results of spin dynamics reported in three different samples. Concluding remarks are discussed in section 4.

## 2. Materials and experiments

To explore the spin dynamics, we investigated here three different samples (namely A, B, and C) grown by MBE on a (001)-oriented GaAs substrate. All samples are symmetrically delta-doped beyond the metal-insulator transition (MIT) where the DP spin relaxation has been reported to be more efficient [28, 29, 30]. For all the samples, the density of Si-doping was $2.2 \times 10^{12}$ cm$^{-2}$ separated from the QW by 7 periods of short-period AlAs/GaAs superlattices with 4 AlAs and 8 GaAs monolayers per period. Sample A is a 45-nm-wide GaAs QW. Owing to a large well width and high electron density the electronic system results in a double quantum well (DQW) configuration by forming a soft barrier inside the well due to the Coulomb repulsion of electrons. Sample B studied here is a triple quantum well (TQW) with 22-nm-thick central well separated from the side wells by 2-nm-thick $Al_{0.3}Ga_{0.7}As$ barriers. Both side wells have an equal width of 10-nm. Sample C is a wide TQW having the same structure of sample B with a barrier thickness of about 1.4 times thinner than that of sample B. It contains a 26-nm-thick central well and two 12-nm-thick lateral wells. For both the TQW samples, the central well width is kept wider than the lateral wells to be populated because, due to the electron repulsion and confinement, the electron density tends to concentrate mostly in the side wells. The estimated density in the side wells is 35 % larger than that in the central well. The characteristics of the studied samples are summarized in table 1.

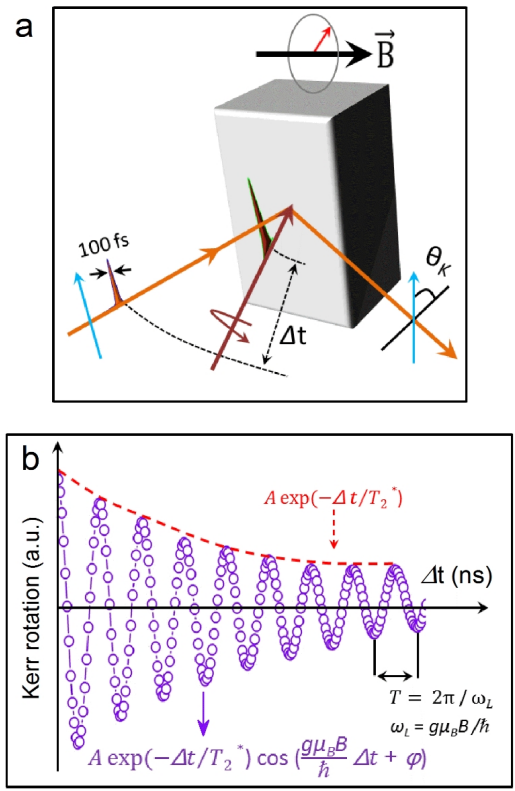

**Figure 1**. (a) Schematic of time-resolved pump-probe technique. The spin polarization are generated by the circularly polarized pump and detected by a time-delayed weak linearly polarized probe pulse. (b) Typical Kerr rotation signal as a function of time delay between pump and probe pulses.

We employed time-resolved pump-probe Kerr rotation (KR) [31] and resonant spin amplification (RSA) [32] techniques to monitor the spin precession of 2DEGs confined in multilayer structures. A Ti:sapphire laser, with 100 fs pulses and repetition frequency ($f_{rep}$) of 50 kHz was used for optical excitation. The light beam was split into the pump and probe beams by a beam splitter. Spin polarization along the structure growth direction were generated by focusing the circularly polarized pump pulses at nearly normal incidence to ap-proximately 50 $\mu$m on the sample surface. The evolution of those optically generated spin ensemble can be monitored via rotation of polarization plane of a linearly polarized probe pulse reflected by the sample. It is accomplished with the help of a mechanical delay line that varies the optical path length of one beam relative to the other. An external magnetic field $B_{ext}$ is applied perpendicular to the structure growth direction (Voigt geometry) as shown in Fig. 1(a). The external magnetic field force the spin precesses around it. The amount of polarization rotation ($\Theta_K$) of the probe beam upon the reflection on the sample is a direct measure of the amount of spin orientation at that moment. This small rotation of the direction of linear polarization can be detected using the balanced bridge. A typical oscillatory response of such a TRKR experiment in the presence of an applied external magnetic field is shown in Fig. 1(b). The frequency of oscillations is a direct measure of electron g-factor, $g = \hbar\omega_L/\mu_B B_{ext}$, while the exponential decay envelope gives a spin dephasing time $T_2^*$. The combination of both, spin dephasing and spin precession leads to an exponentially decaying cosine function of the Kerr rotation described by:

$$\Theta_K = A \exp\left(\frac{-\Delta t}{T_2^*}\right) \cos\left(\frac{|g|\mu_B B_{ext}}{\hbar}\Delta t + \varphi\right) \quad (1)$$

**Table 1**. Studied structures where $n_s$ and $\mu$ are the total electron density and mobility in the QW, determined by electrical transport measurements at low temperature, respectively.

| Name | Structure | QW width (nm) | $n_s$ (cm$^{-2}$) | $\mu$ (cm$^2$/Vs) |
|---|---|---|---|---|
| Sample A | DQW | 45 | $9.2 \times 10^{11}$ | $1.9 \times 10^6$ |
| Sample B | TQW | 10-22-10 | $9.0 \times 10^{11}$ | $5.0 \times 10^5$ |
| Sample C | TQW | 12-26-12 | $9.6 \times 10^{11}$ | $5.5 \times 10^5$ |

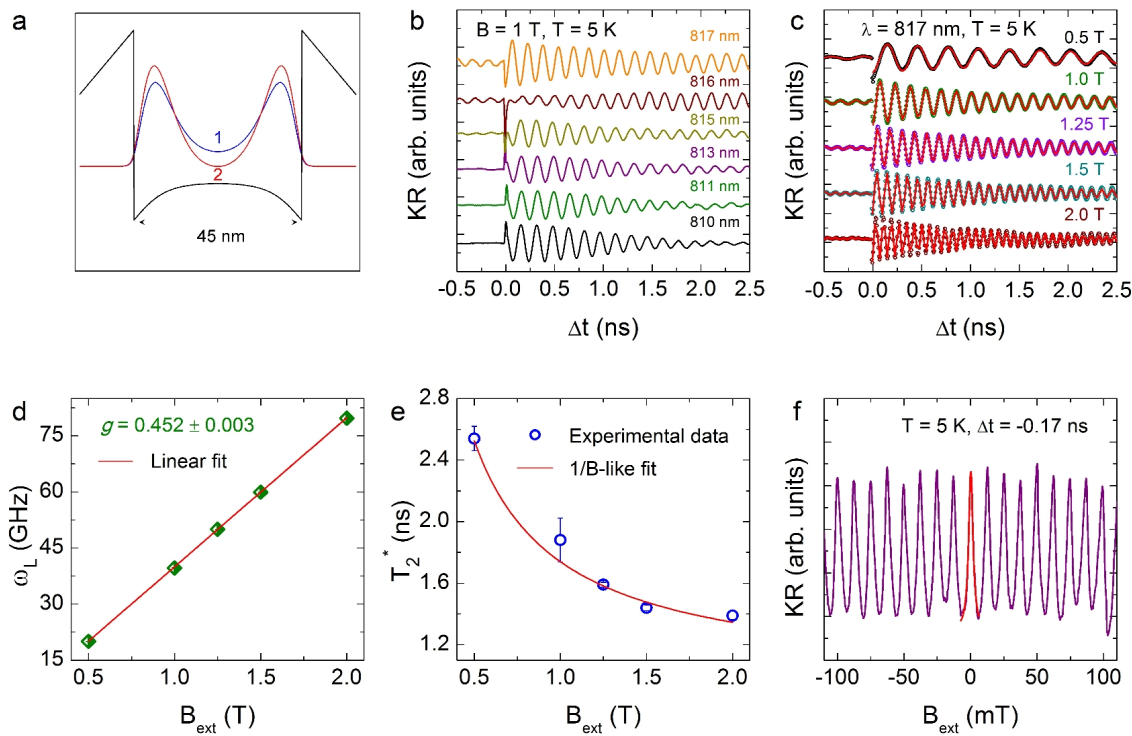

**Figure 2**. Spin dynamics in Sample A: (a) DQW band structure and charge density for the two occupied subbands. (b) TRKR traces measured at $T = 5$ K for different excitation wavelengths. (c) TRKR responses measured at $\lambda = 817$ nm for different magnetic fields. Experimental traces are shown by symbols while the solid lines represent the fitted curves using Eq. 1 (d) $\omega_L$ and (e) $T_2^*$ retrieved from the fit as a function of $B_{ext}$. (f) RSA signal measured at $\Delta t = -0.17$ ns. The experimental parameters are given inside corresponding panels.

Where $A$ is the initial amplitude, $\mu_B$ is the Bohr magneton, $B_{ext}$ is the external magnetic field, $\hbar$ is the reduced Planck constant, $|g|$ is the electron g-factor, $\varphi$ is the initial phase, and $T_2^*$ is the ensemble dephasing time. The cosine factor reflects spin precession about the external magnetic field.

## 3. Results and Discussion

### 3.1 *Spin dynamics in sample A*

Fig. 2(a) depicts the calculated DQW band structure and charge density for the two closely spaced populated subbands with separation energy $\Delta_{12} = 1.4$ meV and equal subband density $n_s$ [33]. To find the maximum Kerr signal with long dephasing time the TRKR measurement was carried out for different pump-probe wavelengths. The time evolution of Kerr signal for the DQW as a function of excitation wavelengths is shown in Fig. 2(b). For clarity of presentation, the TRKR traces are vertically shifted and normalized to $\Delta t = 0$. At higher wavelengths, the decay of spin beat is very slow, and the electron spin polarization doesn't completely decay during the pulse repetition period ($t_{rep}=$ 13.2 ns) as a result one can evidence strong negative delay oscillation. Because of the maximum signal at λ = 817 nm, the influence of spin dynamics on the external magnetic field was studied keeping this wavelength for the following discussion. Fig. 2(c) displays the pump-probe delay scan of KR signal recorded at $T$ = 5 K for various magnetic fields with pump/probe power of 1 mW/300 μW. In the presence of an applied magnetic field, the TRKR signal results in weakly damping oscillations.

To get $T_2^*$ and electron *g*-factor, the TRKR traces were fitted (red curves) to Eq: 1. The fitted values of $\omega_L$ (half-filled diamonds) and $T_2^*$ (open circles) are plotted, in Fig 2(d) and (e), as a function of $B_{ext}$. The precession frequency increases with magnetic field, where, the linear interpolation of the data, shown by solid red line, yields $|g|$ = 0.452 ± 0.003 which is close to the published value of $|g|$ = 0.44 for the bulk GaAs [34] and similar to the one reported for quasi-two-dimensional system with two occupied subbands [35]. $T_2^*$ decreases with growing magnetic field due to the inhomogeneous spread of ensemble *g*-factor [10] and the DP spin relaxation mechanism [27, 36]. The observed $T_2^*$, being limited by variation in the electron *g*-factor Δ*g*, follows 1/*B*-like behavior. Data analysis allows us to estimate the size of this dispersion in *g*-factor by fitting the data to $1/T_2^*(B) = \Delta g \mu_B B / \sqrt{2}\hbar$ [10]. Such a fit to the data, shown by solid line, yields Δ*g* = 0.002 which is 0.41 % of measured *g*-value.

The observation of spin precession at negative Δ*t* indicates that those signal lasts from the previous pump pulse which overlaps with the signal from the following pulse and hence complicates the evaluation of $T_2^*$. In such situations, the RSA technique, based on the interference of spin polarizations generated by subsequent pulses can be used to retrieve the accurate value of $T_2^*$. Fig. 2(f) shows the KR traces obtained by scanning $B_{ext}$ in the range from -100 mT to +100 mT while keeping Δ*t* fixed at -0.17 ns. We observed a series of Lorentzian resonance peaks with spacing $\Delta B = hf_{rep}/g\mu_B \sim 12.5$ mT. The line width of those resonance peaks allows to evaluate $T_2^*$ by using Lorentzian model [32]:

$$\Theta_K = A/\left[(\omega_L T_2^*)^2 + 1\right] \quad (2)$$

where $T_2^* = \hbar/g\mu_B B_{1/2}$ with half-width $B_{1/2}$. The fitting yields $T_2^*$ = 6.44 ± 0.19 ns which is among the longest $T_2^*$ observed for structures with similar doping levels [30, 37]. Based on previous literature [38, 39, 40], the observed RSA signal corresponds to the regime of isotropic spin relaxation where all the peaks have the same height, and spin components of carriers oriented along the growth axis and normal to it relax at the same rate. In the opposite case, in anisotropic spin relaxation, the spin components of carriers relax at a different rate as a result one can see its influences on the relative amplitudes of RSA peaks.

### 3.2 *Spin dynamics in sample B*

The calculated band structure and charge density of the symmetric triple quantum well (Sample B) is shown in Fig. 3(a). The thin barriers separating the wells lead to the strong tunneling of electron states into different wells. As a result, there are three populated subbands (*i*, *j* = 1, 2, 3) with corresponding energy gapes $\Delta_{12}$ = 1.0 meV, $\Delta_{23}$ = 2.4 meV, and $\Delta_{13}$ = 3.4 meV, obtained from the self-consistent Hartree-Fock calculation, which are in complete agreement with periodicity of the magneto-intersubband (MIS) oscillations [41, 42]. These energy gaps characterize the coupling strength between the wells. To select the right excitation energy for this sample, we first measured KR signal vs Δ*t* for different pump-probe wavelengths [see Fig. 3(b)]. From the experimental traces, it is clearly evident, that at lower wavelengths up to 818 nm the signal displaying a rapidly damping initial part transforming into a slowly decaying oscillatory tail. However, at a higher wavelength the signal last longer than $t_{rep}$ demonstrating that in this structure the signal has maximum intensity when excitation energy is tuned to a higher wavelength.

Panel 3(c) shows the TRKR traces measured at λ = 821 nm for various magnetic field in the range from 0.4 to 2 T. According to previous literature, in highly-doped QWs the hole contribution to the electron spin dynamics can be found as a shift of the center of gravity of the electron spin precession [10, 11, 43]. In our structure, we found such a contribution at $B_{ext}$ = 2 T as marked by the arrow in Fig. 3(c). The magnetic field dependencies of $\omega_L$ and $T_2^*$ are shown in Fig 3(d). The observed linear dependence of $\omega_L$ on the magnetic field gives a *g*-value of $|g|$ = 0.453 ± 0.002. From the $B_{ext}$ dependence of $T_2^*$, one can witness a strong reduction in $T_2^*$ with growing field, leading to a 1/*B*-like dependence. The observed dependence assume Δ*g* = 0.0005 (0.10 % of obtained *g*-value). To see the influence of sample temperature on the electron spin dynamics, the delay scan of KR signal was carried out at three different temperatures. Obviously, the signal lasts longer at low temperature as reflected by strong negatively delay oscillations. Additionally, the signal is robust against temperature and was traced up to 250 K.

To avoid the contribution of variation in the ensemble *g*-factor to the spin relaxation process, the spin dynamics presented in Fig. 3(f) was measured in the limit of lowest possible magnetic fields. For that we used the RSA technique by scanning $B_{ext}$ over a range of -150 mT to 150 mT while keeping the delay time fixed at Δ*t* = -0.24 ns. Fitting the zero-field RSA max-imum, using the Lorentzian model, leads to the out-of-plane dephasing time of 13.6 ± 2.07 ns. Apart from the long-lived spin coherence, we observed a strong spin relaxation anisotropy between the electron spins oriented in-plane and out-of-plane as apparent from the suppression of zeroth-field peak compared to the side peaks. The observed anisotropy has its origin in the presence of an internal magnetic field. The magnitude and direction of this internal field can be inferred by fitting the data to the model described in Ref. [24, 44]. Such a fitting, displayed in a selected range (from -50 mT to 50 mT) for clarity, yields the internal field magnitude of $B_\perp$ = 3 mT. Detailed study of spin relaxation anisotropy as a function of experimental parameters (namely, sample temperature, pump-probe delay and optical power) can be found in Ref. [24]. The observed long $T_2^*$ in the out-of-plane direction for both sample A and B stipulates that the scattering-induced DP mechanism is weak in the studied structures [12]. However, combined with the inhomogeneous spread of g-factor it leads to a strong $T_2^*$ reduction.

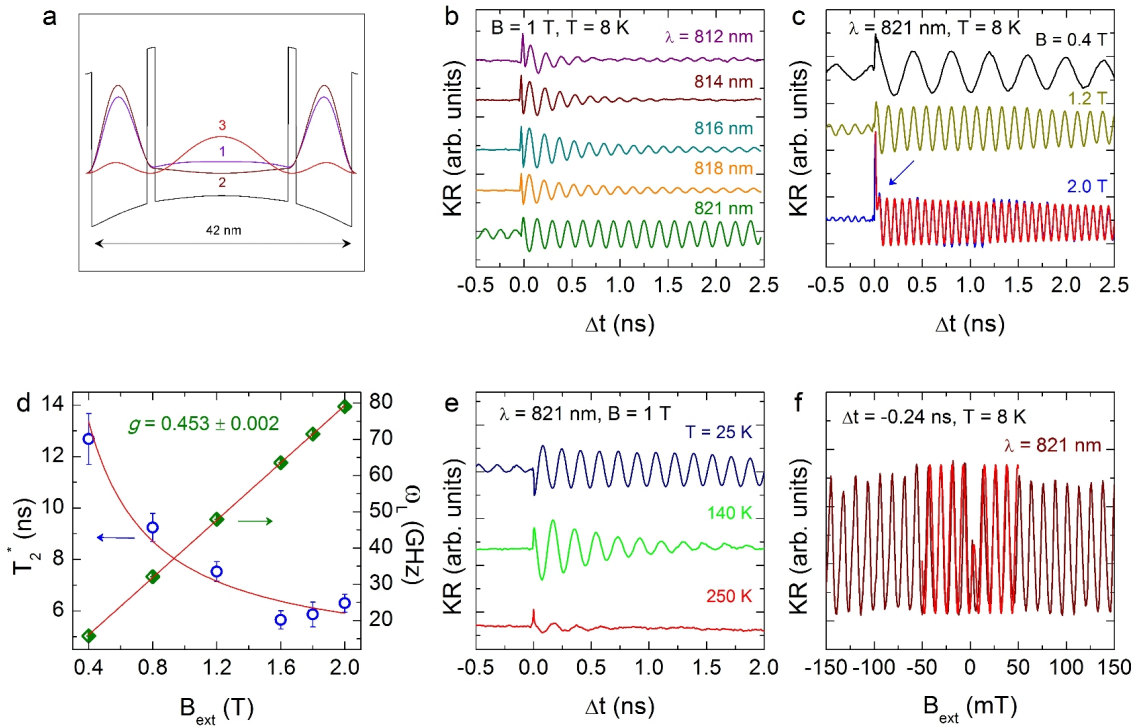

**Figure 3**. Spin dynamics in Sample B: (a) TQW (sample B) band structure and charge density for the three occupied subbands with subband separation $\Delta_{12}$ =1.0 meV and $\Delta_{23}$ = 2.4 meV and $\Delta_{13}$ = 3.4 meV. (b) KR signal measured for different excitation wavelengths at $T$ = 8 K. (c) KR as a function of $\Delta t$ recorded for the different magnetic field at $\lambda$ = 821 nm. The red curve on the top of experimental trace (blue) is a bi-exponential decaying cosine fit to the data. (d) $T_2^*$ and $\omega_L$ as a function of applied magnetic field. (e) TRKR traces recorded at various temperature in the range up to 250 K. (f) RSA scan measured for $\lambda$ = 821 nm.

### 3.3 *Spin dynamics in sample C*

Finally, we report on the spin dynamics for sample C. The band structure and charge density of this structure, with the three populated subbands, is displayed in Fig. 4(a). Contrary to the DQW, we noticed that for both the TQWs the third subband has the opposite charge distribution compared to the first and second subbands. The third subband has charge density localized in the central well, while the electrons in the lower subbands are more distributed in the side wells. The TRKR signals measured by the changing excitation wavelength of the laser pulses over the range from 811 nm to 821 nm, while keeping the same pump power under an external magnetic field of 1 T, are shown in Fig. 4(b). From experimental curves, it is clearly evident that the decay of the Kerr rotation signals is changing with excitation wavelengths that is only for lower wavelengths the spin precession lasting up to $\Delta t$ = 2 ns. To get information over spin dephasing time and electron g-factor the precessional signal was fitted with mono-exponential decaying cosine function as shown by red curves plotted on the top of experimental data. The obtained $T_2^*$ and g-factor are shown in Fig. 4(c). A clear variation of spin dephasing time is observed with the increase of laser detuning having the maximum at 811 nm. Additionally, the electron g-factor shows a variation of 0.028, due to the change in precession frequency as marked by the dashed line, in the measured range of wavelengths.

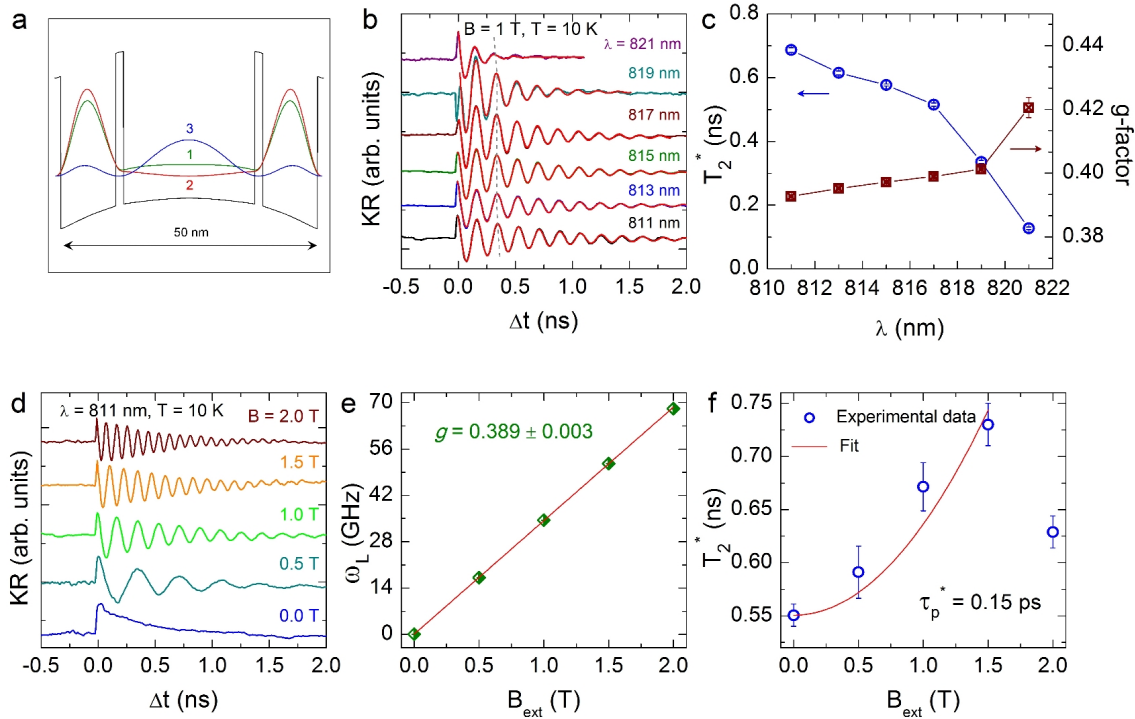

**Figure 4**. Spin dynamics in Sample C: (a) Sample C band structure and charge density for the three occupied subbands. (b) TRKR traces recorded at $T = 10$ K for different pump-probe wavelengths (colored) and fits to the data (red). Where the spin beats live longer at lower wavelengths (c) The relative $T_2^*$ and $g$-factor evaluated from b. (d) Dependence of Kerr signal on external magnetic fields and corresponding (e) $\omega_L$ and (f) $T_2^*$.

To investigate the dependence of spin dynamics on the applied magnetic field a series of TRKR measurements, for the wavelength with maximum KR signal, were performed at $T = 10$ K. Fig. 4(d) shows TRKR scans measured with no magnetic field and in the transverse magnetic field up to 2 T. In the frame of DP spin relaxation mechanism, the observed signal at $B_{ext} = 0$ corresponds to the strong scattering regime [5]. Unlikely, in the weak scattering regime, the electrons spin precess about the spin-orbit field by one or more revolutions before scattering and hence leading to an oscillatory behavior [45]. From TRKR signals the dependence of spin dephasing time and Larmor frequency on the applied magnetic field is plotted in Fig. 4(e & f). The linear dependence of Larmor frequency on applied magnetic field yields an effective Lande factor of 0.389 ± 0.003 which is in agreement with the magnitude of $g$-value reported previously on the same sample [6].

Additionally, with growing magnetic field up to 1.5 T, we observed a monotonous increase in $T_2^*$. In such situation, the $B_{ext}$ leads to the cyclotron motion of conduction band electrons which lets the direction of $\mathbf{k}$ to change, thereby suppressing the precession around the random internal magnetic field. As a result, the electron spin preserves its initial spin orientation, and thus inconsistent with DP mechanism [46], leading to the enhancement of $T_2^*$. This is the key difference compared to sample A and B where cyclotron effect is suppressed, and the DP mechanism is more efficient. The dependence of $T_2^*$ on the applied magnetic field follows a quadratic dependence given as [46, 37]:

$$T_2^*(B) = T_2^*(0)/\left(1 + \omega_c^2 \tau_p^2\right) \quad (3)$$

Here, $T_2^*(0)$ is the zero-field spin relaxation time, $\omega_c$ is the cyclotron frequency, and $\tau_p$ is the single electron relaxation time, which is defined as the inverse sum of transport scattering rate and electron-electron scattering rate [5]

$$\tau_p = \left(\tau^{-1} + \tau_{ee}^{-1}\right)^{-1} \quad (4)$$

From fit (solid red curve) to the data we retrieved $\tau_p = 0.15$ ps which is in agreement with the magnitude of quantum lifetime reported by transport for a similar TQW sample [41]. The transport scattering time was estimated, using the electron charge $e$ and effective mass $m^*$, by $\tau = \mu m^*/e = 20$ ps which is quite different from $\tau_p$. The observed large difference was associated with the fact that $\tau$ includes only the large-angle scattering, while $\tau_p$ is caused by all kind of scattering events. The ratio of measured $\tau$ and $\tau_p$ determine the nature of dominant scattering mechanism [47]. For GaAs based heterostructures, it was assumed that $\tau/\tau_p \lesssim 10$ for background impurity scattering and $\tau/\tau_p \gtrsim 10$ for the remote ionized impurity scattering [47]. The observed $\tau/\tau_p \approx 135$, indicates that the dominant scattering in our structure is caused by remote instead of background impurities. The electron-electron scattering time ($\tau_{ee}$) can be approximated by using Eq. 4 which leads to $\tau_p \approx \tau_{ee}$ demonstrating the supremacy of electron-electron scattering over microscopic scattering mechanisms [48].

## 4. Conclusions and outlook

In summary, we have studied the electron spin dynamics in high-mobility two-dimensional electron gases using the pump-probe reflection techniques: time-resolved Kerr rotation and resonant spin amplification. The DQW structure yields $T_2^* = 6.44$ ns, while, in the TQW we observed a strikingly long $T_2^*$ exceeding the laser repetition. In the wide TQW, $T_2^*$ increases with the applied magnetic field but is much smaller than that in the DQW and other TQW. Additionally, we observed anisotropic feature due most likely to the presence of an internal magnetic field. The observed long-lived spin coherence persists up to about room temperature, with encouraging indication for spin-optoelectronics and particularly the long spin memories in multilayer GaAs QWs. We believe that the determination of all the relevant time scales will be useful for the future spintronics devices and quantum information processing.

## Acknowledgement

F.G.G.H. acknowledges financial support from Grant No. 2009/15007-5, 2013/03450-7, 2014/25981-7 and 2015/16191-5 of the São Paulo Research Foundation (FAPESP). S.U acknowledges TWAS/CNPq for financial support.